\documentclass[pdflatex,sn-mathphys-num]{sn-jnl}

\usepackage{amsmath,amssymb,graphicx,bm}

\newcommand{\beq}{\begin{equation}}
\newcommand{\eeq}{\end{equation}}
\newcommand{\beqa}{\begin{eqnarray}}
\newcommand{\eeqa}{\end{eqnarray}}

\begin{document}

\title{Contrasting GHZ and W-state Entanglement Dynamics due to Correlated Markov Noise}

\author{\fnm{Stephen} \sur{Brockerhoff}}\email{stephen.brockerhoff@cooper.edu}

\author*{\fnm{Brittany} \sur{Corn-Agostini}}\email{brittany.corn@cooper.edu}

\affil{Department of Physics, Albert Nerken School of Engineering, The Cooper Union for the Advancement of Science and Art, New York City, New York 10003, USA}

\date{\today}

\abstract{
The ability to preserve multipartite entanglement in noisy environments is central to advancing quantum information processing. In this work, we develop a semiclassical theoretical model of three entangled qubits exposed to local Markov noise environments with tunable statistical correlations between noise sources. We show that such correlations can significantly influence the dynamics of multipartite entanglement, in some cases slowing its decay and, under ideal conditions, even enabling full preservation. Using tripartite negativity as an entanglement measure, we derive analytical results for the GHZ and W states, demonstrating their contrasting responses to correlated and anticorrelated noise. Our analysis identifies regimes in which multipartite entanglement can be sustained despite environmental interactions, offering new insight into how noise correlations may serve as a resource for protecting quantum coherence in multi-qubit systems.
}

\maketitle

\section{Introduction}
Entanglement is a fundamental feature of quantum mechanics with no classical analogue \cite{Ent1}. It serves as a central resource in quantum information processing, enabling protocols in quantum computing \cite{QComp1}, quantum communication \cite{QComm1}, and quantum cryptography \cite{QCrypt1}. However, in realized scenarios, quantum systems are inevitably coupled to their environment, resulting in decoherence \cite{Decoh1, ExperimentalEntanglementDecoherence1, ExperimentalEntanglementDecoherence2} and the degradation of entanglement \cite{Disentanglement1, Dephasing2, ESD1, ESD2, ESDexperimental}. Understanding the dynamics of entanglement in open quantum systems is therefore essential for the development of scalable quantum technologies \cite{OQS1, OQS2}.

Previous theoretical studies have demonstrated the potential for entanglement modulation and rebirth \cite{Revival1, Revival2, QIP2009} as well as the effects of noise correlations on qubit dynamics \cite{correlated1, correlated2}. In this work, we examine the dynamics of multipartite entanglement in a system of three uncoupled qubits, each subject to local, classical environmental noise, known to inevitably lead to disentanglement \cite{Yusui2014, Past3QWork1, Past3QWork2}. By introducing statistical cross-correlations between the noise sources, we explore how such correlations can mitigate or accelerate entanglement decay. Using tripartite negativity as our entanglement measure, we analyze the time evolution of two fundamental three-qubit states, the GHZ state and the W state. Our results demonstrate distinct responses by these states to correlated versus anticorrelated noise, revealing conditions under which multipartite entanglement can be sustained despite environmental interactions.

\section{Three-Qubit Markov Master Equation}

\begin{figure}[b]
\centering
\includegraphics[height=.28\columnwidth]{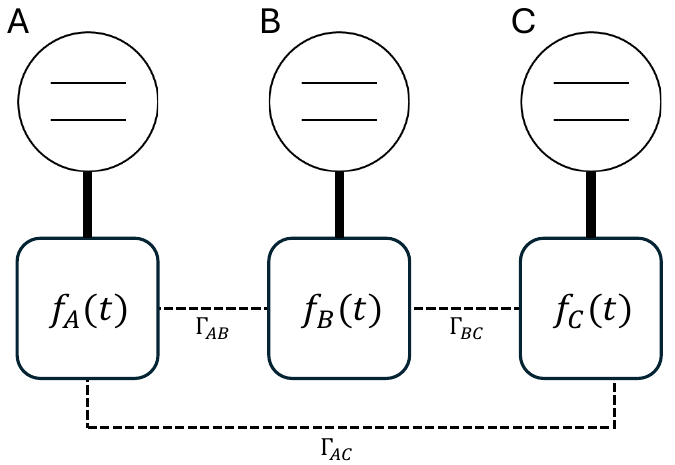}
\caption{Three uncoupled qubits subject to local classical noise that may be statistically correlated.}
\label{modelfigure}
\end{figure}

Our model consists of three identical, uncoupled qubits, labeled $A$, $B$, and $C$, each interacting with its own local noise source modeled as a real-valued stochastic field, $f_A(t)$, $f_B(t)$, and $f_C(t)$, respectively, as depicted in Fig. \ref{modelfigure}. The effective Hamiltonian describing the qubit dynamics (with $\hbar=1$) is \cite{QIP2009}
\beq
\label{Heff}
\hat{H}_{\text{eff}} = \hat{H}_{\text{sys}} + f_A(t)\hat{\sigma}_A + f_B(t)\hat{\sigma}_B + f_C(t)\hat{\sigma}_C,
\eeq
where the system Hamiltonian is
\beq
\hat{H}_{\text{sys}} = \frac{\omega}{2} \left( \sigma_z^A + \sigma_z^B + \sigma_z^C \right).
\eeq

For simplicity, we take the qubits to be identical, each with transition frequency $\omega$, and assert that each qubit interacts only with its corresponding local noise field. We consider two distinct types of noise channels: (i) dephasing noise, $\hat{\sigma}_i = \sigma_z^i$, which suppresses phase coherence without energy loss \cite{Dephasing1, Dephasing2}, and (ii) amplitude noise, $\hat{\sigma}_i = \sigma_x^i$, which induces dissipative dynamics and can lead to sudden entanglement death \cite{ESD1, ESD2}  

The noise fields are assumed to be Gaussian white noise with the following statistical properties for $i,j \in \{A,B,C\}$ and $i \neq j$:
\beqa
M[f_i(t)] &=& 0, \\
M[f_i(t) f_i(s)] &=& \tfrac{\gamma_i}{2}\,\delta(t-s), \\
M[f_i(t) f_j(s)] &=& \tfrac{\Gamma_{ij}}{2}\,\delta(t-s).
\eeqa

Here, $M[\cdot]$ denotes the ensemble average, and the Dirac delta $\delta(t-s)$ enforces the Markovian character of the noise. The coefficients $\gamma_i$ specify the local decoherence (dephasing or dissipation) rates of the qubits. For simplicity, we set $\gamma_A = \gamma_B = \gamma_C \equiv \gamma$. The parameters $\Gamma_{ij}$ quantify the degree of cross-correlation between pairs of noise environments, subject to the constraint $\Gamma_{ij} \leq \gamma$. When $\Gamma_{ij} = \gamma$, the corresponding noise sources are perfectly correlated, effectively mimicking a common environment; when $\Gamma_{ij} < \gamma$, they are only partially correlated. Further constraints on the allowed values of $\Gamma_{ij}$ will be discussed in Sec. III.  

The Markov master equation governing the time evolution of the reduced density matrix of the three-qubit system, $\rho(t)$, is derived from the stochastic Schrodinger equation \cite{GisinQSD, DiosiSSE} governed by $\hat{H}_\text{eff}$:
\beqa
\label{master_eq}
\dot{\rho}= -i[\hat{H}_{\text{sys}}, \rho] &-&\gamma_A(\rho - \hat{\sigma}_A\rho\hat{\sigma}_A) -\gamma_B(\rho - \hat{\sigma}_B\rho\hat{\sigma}_B) -\gamma_C(\rho - \hat{\sigma}_C\rho\hat{\sigma}_C) \\
&-&\Gamma_{AB}(\hat{\sigma}_A\hat{\sigma}_B\rho + \rho\hat{\sigma}_A\hat{\sigma}_B - \hat{\sigma}_A\rho\hat{\sigma}_B - \hat{\sigma}_B\rho\hat{\sigma}_A) \nonumber \\
&-&\Gamma_{AC}(\hat{\sigma}_A\hat{\sigma}_C\rho + \rho\hat{\sigma}_A\hat{\sigma}_C 
- \hat{\sigma}_A\rho\hat{\sigma}_C - \hat{\sigma}_C\rho\hat{\sigma}_A) \nonumber \\
&-&\Gamma_{BC}(\hat{\sigma}_B\hat{\sigma}_C\rho + \rho\hat{\sigma}_B\hat{\sigma}_C 
- \hat{\sigma}_B\rho\hat{\sigma}_C - \hat{\sigma}_C\rho\hat{\sigma}_B) \nonumber
\eeqa
(where, by convention, $\hbar=1$). The first line of Eq. \ref{master_eq} reflects the Lindblad form of the master equation \cite{Lindblad} of three qubits interacting only with local dephasing or dissipative noise, while the last three terms harness the impact of correlated noise, to be analyzed in the following sections. 

Without loss of generality, we present the Markov master equation for any number of uncoupled qubits in the presence of local noise environments having varying strengths of cross-correlation:
\beqa
\dot{\rho}=-i[\hat{H}_{sys}, \rho] &-& \sum_j \gamma_j(\rho - \hat{\sigma}_j\rho\hat{\sigma}_j) \\ 
&-& \sum_{j}\sum_{k<j} \Gamma_{jk} (\hat{\sigma}_j \hat{\sigma}_k \rho + \rho \hat{\sigma}_j \hat{\sigma}_k - \hat{\sigma}_j \rho \hat{\sigma}_k - \hat{\sigma}_k \rho \hat{\sigma}_j) \nonumber
\eeqa

\section{Noise Correlation Strength}
While each pairwise cross-correlation is bounded by $\tfrac{\Gamma_{ij}}{\gamma} \leq 1$, the three noise variables are not independent. In order for the full correlation matrix to remain positive semi-definite, the following constraint must be satisfied:
\begin{equation}
    \gamma^3 - \gamma\Gamma_{AB}^2 - \gamma\Gamma_{AC}^2 - \gamma\Gamma_{BC}^2 + 2\Gamma_{AB}\Gamma_{BC}\Gamma_{AC} \ge 0.
\end{equation}

\begin{figure}[h]
\centering
\includegraphics[width = .6\columnwidth]{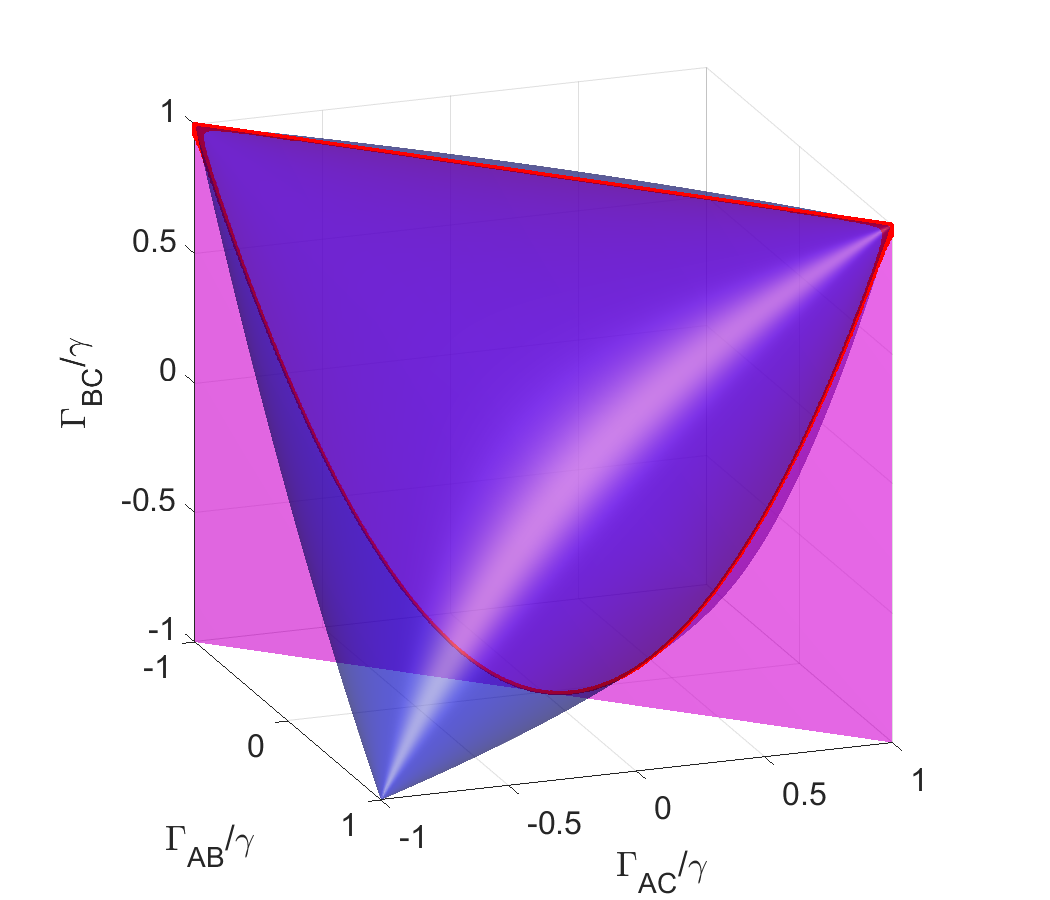}
\caption{The surface of the cubic volume defined in Equation (8) is shown in blue. The plane $\Gamma_{AB}=\Gamma_{AC}$ is shown in pink. Their intersection, the path 'PQRS', is shown in red. } 
\label{blob}
\end{figure}

This inequality defines a three-dimensional volume in correlation space, parameterized by the coordinates $\big(\tfrac{\Gamma_{AB}}{\gamma}, \tfrac{\Gamma_{AC}}{\gamma}, \tfrac{\Gamma_{BC}}{\gamma}\big)$, as shown in Fig. \ref{blob}. Of particular interest is the intersection of the surface with the plane $\Gamma_{AB}=\Gamma_{AC}$, which yields the red curve we will refer to as path "PQRS". Along this path, several special points correspond to physically meaningful correlation limits:
\begin{table}[h!]
\begin{tabular}{@{}lp{0.70\textwidth}@{}}
P(0,0,-1) & Environments of qubit pairs AB and BC are uncorrelated while that of qubit pair AC are maximally anti-correlated; \\[8pt]
Q(1,1,1) & Environments of all qubit pairs are maximally correlated; \\[8pt]
R(-1, -1, 1) & Environments of qubit pairs AB and BC are maximally anti-correlated and therefore environments of qubit pair BC are maximally correlated; \\[8pt]
S(-$\frac{1}{2}$, -$\frac{1}{2}$, -$\frac{1}{2}$) & Environments of all qubit pairs are equally partially anti-correlated.
\end{tabular}
\end{table}

By analyzing different correlation strengths between pairs of qubit environments, we will show how correlated noise influences the evolution of tripartite entanglement. 

\section{Results}

In the following sections, we present the results obtained by solving the master equation in Eq.~\ref{master_eq} for two types of local noise: dephasing noise, $\hat{\sigma}_i = \sigma_z^i$, and amplitude noise, $\hat{\sigma}_i = \sigma_x^i$. Multipartite entanglement for the three-qubit system is quantified using the tripartite negativity \cite{TripartiteNegativity, EntanglementMeasureReview, Negativity},
\beq
\mathcal{N}_{ABC} = \sqrt[3]{\mathcal{N}_{A|BC}\,\mathcal{N}_{B|AC}\,\mathcal{N}_{C|AB}},
\eeq
where $\mathcal{N}_{i|jk}$ is the negativity of bipartite divisions $i|jk$, defined as
\beq
\mathcal{N}_{i|jk} = \|\rho^{T_i}\|_1 - 1.
\eeq
Here $\rho^{T_i}$ is the partial transpose with respect to qubit $i$ and $\|\cdot\|_1$ is the trace norm.  

As representative initial states, we consider the Greenberger–Horne–Zeilinger (GHZ) state,  
\beq
\lvert GHZ \rangle = \tfrac{1}{\sqrt{2}}(\lvert 000 \rangle + \lvert 111 \rangle) \nonumber
\eeq
with $\mathcal{N}_{GHZ}=1$, and the $W$ state,  
\beq
\lvert W \rangle = \tfrac{1}{\sqrt{3}}(\lvert 001 \rangle + \lvert 010 \rangle + \lvert 100 \rangle) \nonumber
\eeq
with $\mathcal{N}_{W}=\tfrac{2\sqrt{2}}{3}\approx 0.94$.  

By systematically varying the degree of cross-correlations between the noise environments, we uncover contrasting behaviors of these two canonical three-qubit states. In particular, we show how correlated and anticorrelated noise can either suppress or preserve multipartite entanglement, identifying conditions under which the decay of entanglement can be slowed, or even halted, despite environmental interactions. 

\subsection{Dephasing Noise}

Dephasing noise is both physically relevant in many experimental platforms \cite{Dephasing1, Dephasing2} and analytically tractable; this tractability allows closed-form expressions for the entanglement dynamics of our three-qubit system.

\begin{figure}[h]
\centering
\includegraphics[width=0.6\columnwidth]{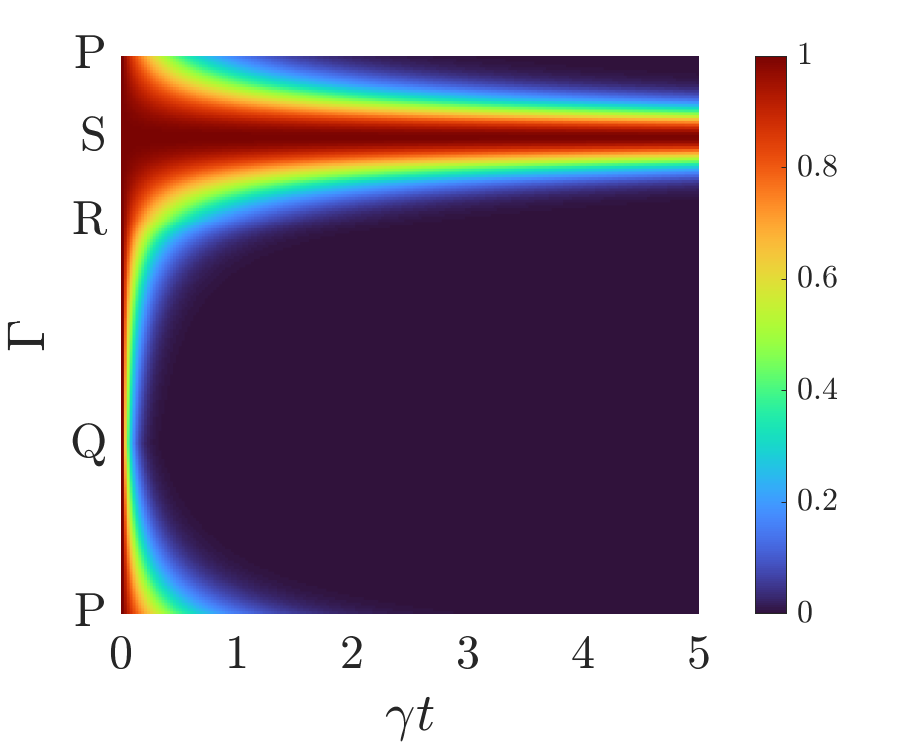}
\caption{Tripartite negativity \(\mathcal{N}_{ABC}\) for the GHZ state under dephasing noise, plotted along the path PQRS. The GHZ state resists disentanglement most strongly at point $S(-\tfrac{1}{2},-\tfrac{1}{2},-\tfrac{1}{2})$, where the noise environments are equally anticorrelated.}
\label{DephGHZ}
\end{figure}

By symmetry of the GHZ state, every bipartition, $i|jk$, has identical negativity, therefore reducing the tripartite negativity to a single function:
\beq
\mathcal{N}_{GHZ}=\mathcal{N}_{A|BC}=\mathcal{N}_{B|AC}=\mathcal{N}_{C|AB}= e^{-6\gamma t -4(\Gamma_{AB}+\Gamma_{AC}+\Gamma_{BC})t}.
\eeq
This result shows the expected exponential decay of entanglement \cite{Dephasing2} at a base rate of $6\gamma$ from the decoherence of each qubit in its local environment. An additional effective contribution then arises from the noise correlations, $\Gamma_{\text{eff}}=\Gamma_{AB}+\Gamma_{AC}+\Gamma_{BC}$, that either accelerates or decelerates disentanglement. Most notably, when \(\Gamma_{\text{eff}}=-\tfrac{3}{2}\gamma\), its lower bound occurring uniquely at point $S(-\tfrac{1}{2},-\tfrac{1}{2},-\tfrac{1}{2})$ in Figure \ref{DephGHZ}, the exponential factor vanishes and disentanglement is fully suppressed: anticorrelated dephasing Markov noise can thus, in this ideal limit, preserve GHZ entanglement. While positive correlations increase the effective dephasing rate and accelerate entanglement loss, as shown in Figure 3 in the vicinity of point $Q(1, 1, 1)$, statistical anticorrelations reduce the effective decoherence experienced by the GHZ state and can significantly slow disentanglement.

The W state behaves qualitatively differently from the GHZ state under noise correlations. Without correlations, this difference is not quite apparent, as the tripartite negativity of the W state under local dephasing Markov noise also simply decays exponentially, with $\mathcal{N}_{W} = \frac{2\sqrt{2}}{3} e^{- 4\gamma t}$. However, once introducing pairwise correlations, a break in the bipartition entanglement symmetry of the W state is revealed: the negativity for the bipartition $i|jk$ becomes
\beq
\mathcal{N}_{i|jk}=\frac{2}{3} e^{-4\gamma t}\sqrt{e^{8 \Gamma_{ij} t}+e^{8 \Gamma_{ik} t}}.
\eeq
This expression quantifies the mixedness of qubit $i$ with the subsystem $jk$ and depends explicitly on the pairwise correlations between $i$'s environment and those of $j$ and $k$. Combining the three bipartitions yields the tripartite negativity

\beq
\mathcal{N}_W = \frac{2}{3} e^{-4\gamma t} \prod_{\substack{i,j,k\in\{A,B,C\} \\ i \neq j \neq k}} \big(e^{8\Gamma_{ij}t}+e^{8\Gamma_{i k}t}\big)^{1/6},
\eeq
where the product cycles over the three distinct bipartitions.

\begin{figure}[h]
\centering
\includegraphics[width=0.6\columnwidth]{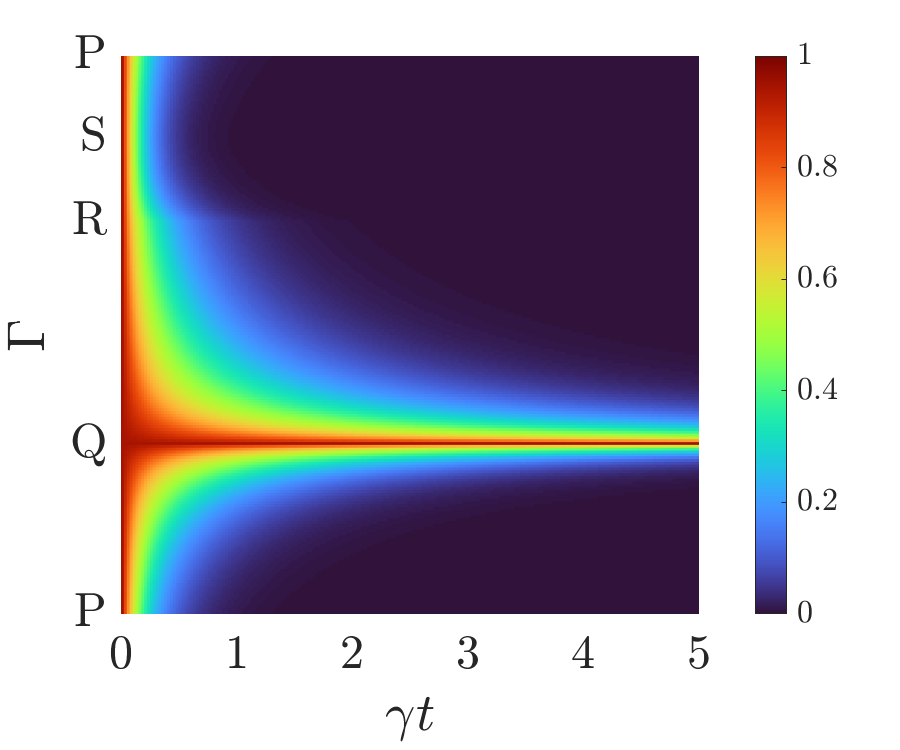}
\caption{Tripartite negativity \(\mathcal{N}_{ABC}\) for the W state under dephasing noise, plotted along the path PQRS. W state entanglement is most robust at point $Q(1, 1, 1)$, where all noise environments are maximally correlated.}
\label{DephW}
\end{figure}

Unlike the GHZ state, which exhibits a
summative effect from cross-correlated environments, $\Gamma_{\text{eff}}$, the W-state dependence on noise correlations is product-like: each bipartition is influenced by the pairwise correlations that involve the partitioned qubit. As a consequence, positive (i.e., correlated) pairwise noise tends to protect W-state entanglement. Full preservation occurs in the limit of all-to-all maximal correlation, depicted at point $Q(1,1,1)$ in Figure \ref{DephW}.

\begin{figure}[h]
\centering
\includegraphics[width=.8\columnwidth]{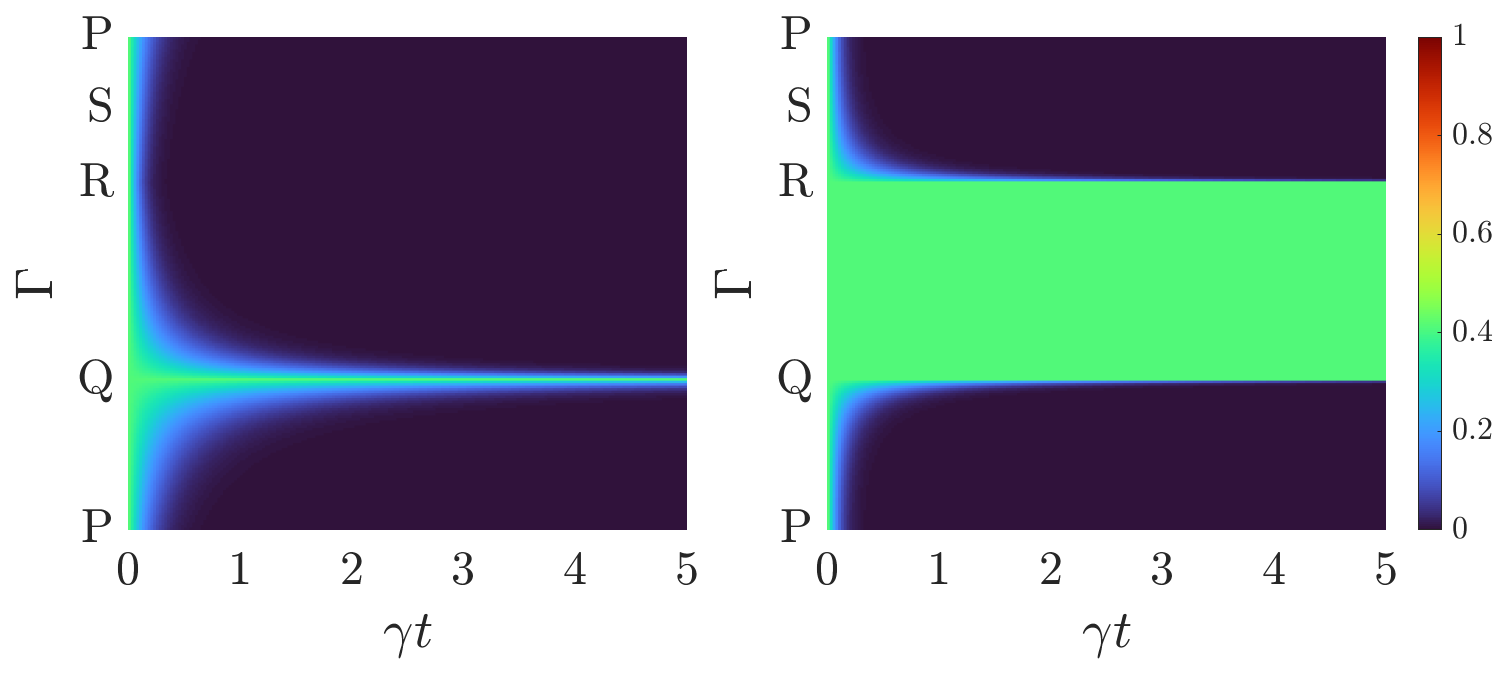}
\caption{Pairwise entanglement dynamics $\mathcal{N}_{A|B}$ and $\mathcal{N}_{A|C}$ (left) and $\mathcal{N}_{B|C}$ (right) for the W state under dephasing noise along correlation path PQRS. (Note that $\mathcal{N}_{A|C}$ = $\mathcal{N}_{B|C}$ due to the choice of $\Gamma_{AB}=\Gamma_{AC}$).}
\label{pairWdeph}
\end{figure}

The qualitative contrast between GHZ and W dynamics stems from the distinct nature of their entanglement. The GHZ state encodes genuinely global tripartite entanglement with zero bipartite entanglement in reduced two-qubit states, so its resilience or fragility is governed by global (sum-like) correlation effects. By contrast, the W state retains bipartite entanglement when one qubit is traced out \cite{TripartiteNegativity}; consequently, its tripartite dynamics are closely tied to the survival of pairwise entanglement, $\mathcal{N}_{i|j}$, as presented in Figure \ref{pairWdeph}. We illustrate that when the environments of a given qubit pair are fully correlated, the corresponding pairwise entanglement is preserved—and since tripartite negativity requires all three bipartitions to be robust, the W state is preserved only when pairwise entanglement is sustained across all pairs.

In summary, for dephasing noise the GHZ and W states respond in opposite ways to the sign of environmental correlations: the GHZ state is favored by anticorrelated environments (sum-like dependence), while the W state is favored by correlated environments (pairwise/product-like dependence). This distinction originates in the GHZ state's purely global entanglement versus the W state's retention of bipartite entanglement in reduced subsystems.

\subsection{Amplitude Noise}

Amplitude noise directly perturbs qubit populations and is among the most destructive decoherence mechanisms in open quantum systems \cite{ESD1, ESD2}. It can drive entanglement to vanish abruptly in finite time, a phenomenon known as entanglement sudden death (ESD) \cite{SuddenDeath}. Unlike the dephasing noise case, the master equation in Eq. \ref{master_eq} becomes analytically intractable once amplitude noise is introduced. Consequently, for amplitude noise, the master equation is solved numerically.

\begin{figure}[h]
\centering
\includegraphics[width=0.49\columnwidth]{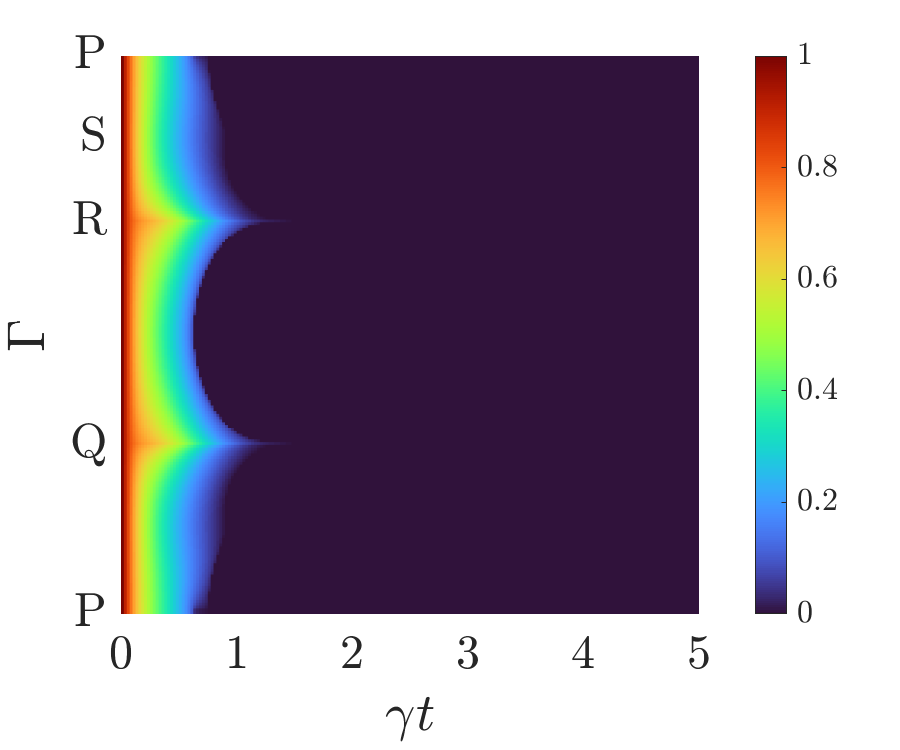}
\hfill
\includegraphics[width=0.49\columnwidth]{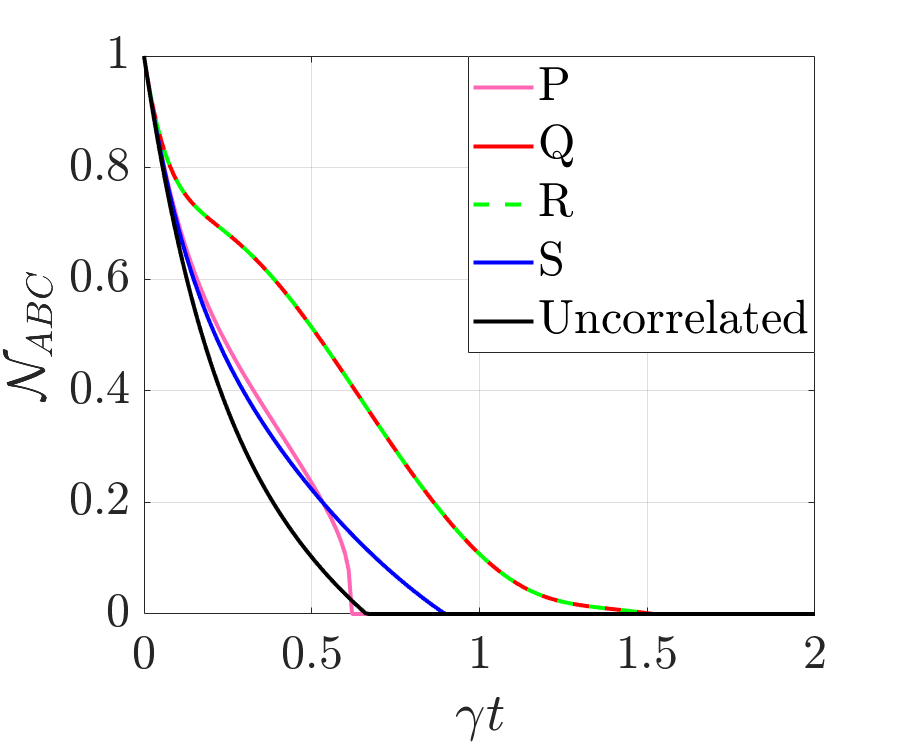}
\caption{(a) Tripartite negativity \(\mathcal{N}_{ABC}\) for the GHZ state under amplitude noise, plotted along the path PQRS. ESD always occurs for the GHZ state under amplitude noise along path PQRS, though multipartite entanglement persists slightly longer at points Q and R. (b) Tripartite negativity \(\mathcal{N}_{ABC}\) plotted from t=0 to t=$2 \gamma^{-1}$ explicitly at points P, Q, R, S, and (0,0,0).}
\label{DissGHZ}
\end{figure}

Figures \ref{DissGHZ}a and \ref{DissGHZ}b show the time evolution of the tripartite negativity for the GHZ state along the correlation path PQRS and at representative points, respectively. Across the entire path, ESD occurs and the tripartite negativity reaches zero at finite time for all correlation choices considered. Nevertheless, the decay rate and approach to zero still depend on the correlation structure. In particular, at $Q(1, 1, 1)$ and $R(-1, -1, 1)$, disentanglement is noticeably delayed relative to the uncorrelated case $(0, 0, 0)$, and the decay becomes smoother as $t$ approaches the ESD time. Thus, while correlations cannot prevent ESD for the GHZ state under amplitude noise, specific correlation patterns tend to slow the loss of tripartite entanglement.

\begin{figure}[h]
\centering
\includegraphics[width=0.49\columnwidth]{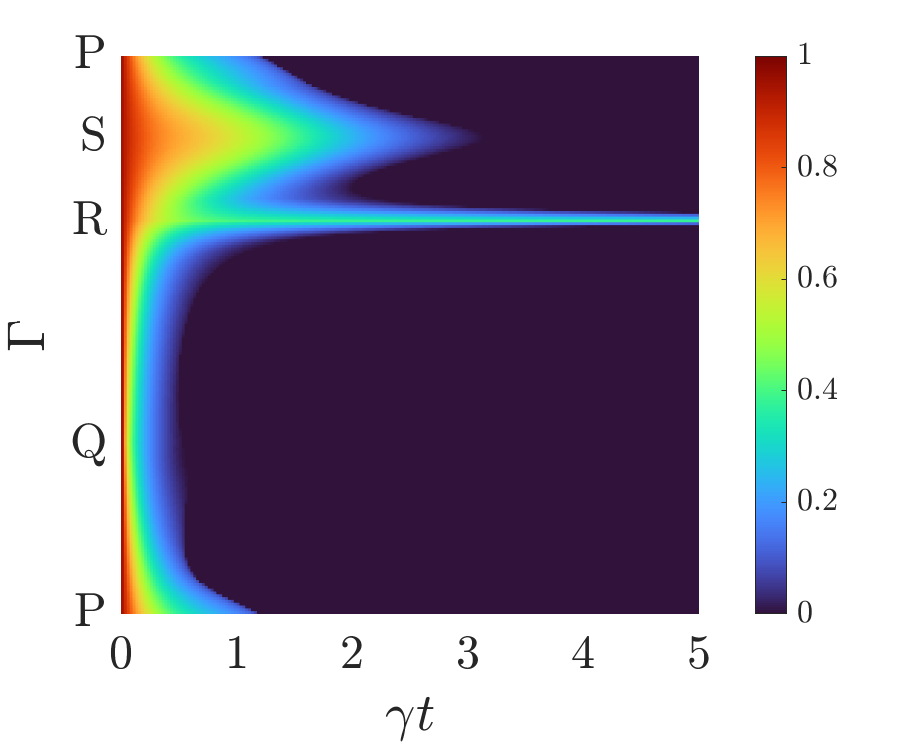}
\hfill
\includegraphics[width=0.49\columnwidth]{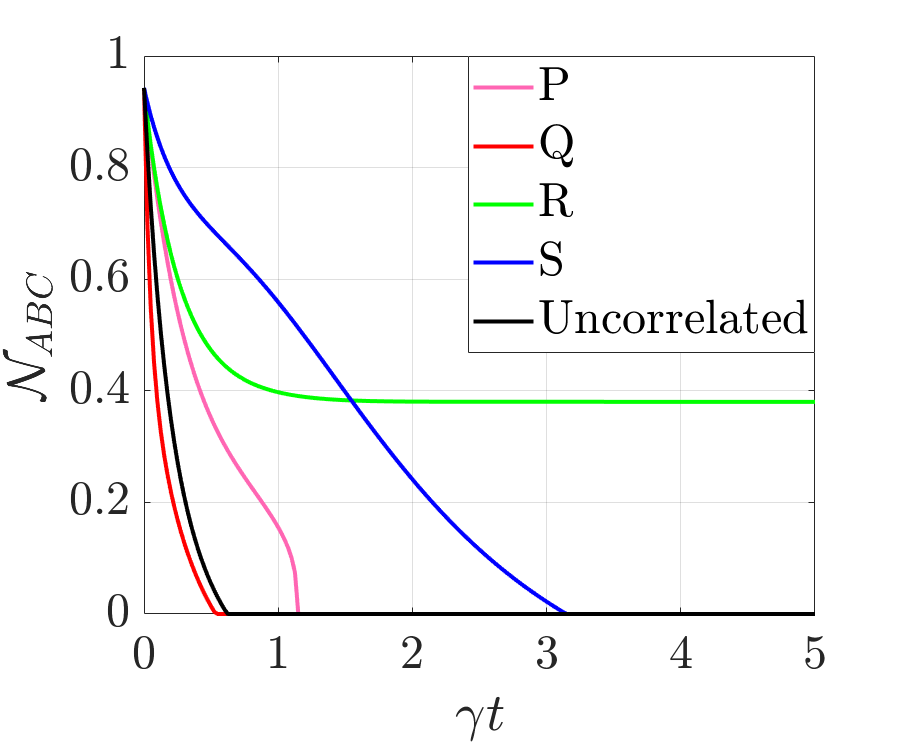}
\caption{(a) Tripartite negativity \(\mathcal{N}_{ABC}\) for the W state under amplitude noise, plotted along the path PQRS. ESD occurs for the W state under amplitude noise at all points on the path PQRS except for point R(-1, -1, 1), where multipartite entanglement is partially preserved. (b) Tripartite negativity \(\mathcal{N}_{ABC}\) plotted from t=0 to t=$5 \gamma^{-1}$ explicitly at points P, Q, R, S, and (0,0,0).}
\label{DissW}
\end{figure}

Figures \ref{DissW}a and \ref{DissW}b illustrate the time evolution of the W state under amplitude noise along the same correlation path PQRS, where the W state exhibits richer features than the GHZ state. Here ESD is observed for all correlation configurations except at $R(-1, -1, 1)$ where the tripartite negativity decays to a finite nonzero asymptote $\mathcal{N}_{ABC}(t\!\to\!\infty)\approx 0.3802$. This residual entanglement arises from the W state's inherent ability to retain bipartite correlations, as was similarly observed in the dephasing case. 

\begin{figure}[b]
\centering
\includegraphics[width=0.8\columnwidth]{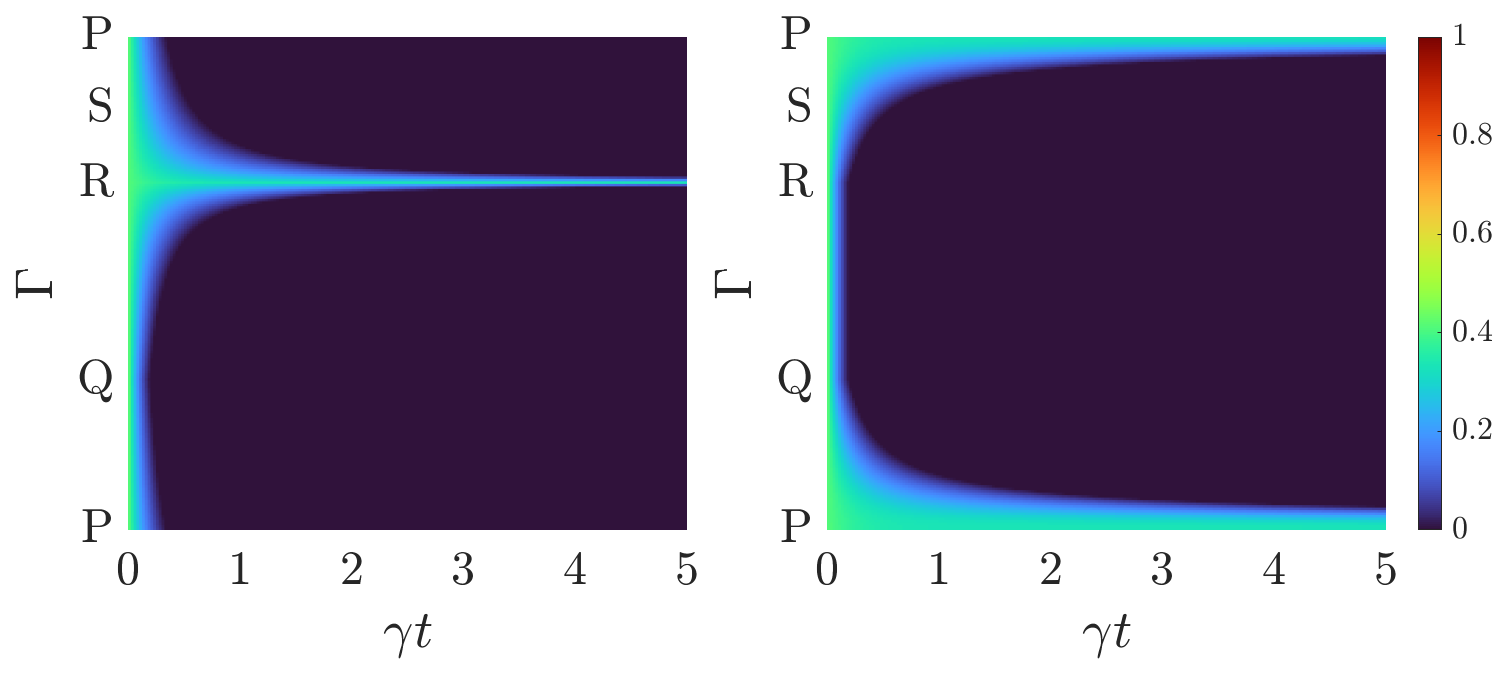}
\caption{Pairwise entanglement dynamics for the W state under amplitude noise along correlation path PQRS for (a) both $\mathcal{N}_{A|B}$ and  $\mathcal{N}_{A|C}$ (left) and $\mathcal{N}_{B|C}$ (right). Note that $\mathcal{N}_{A|B}$ and  $\mathcal{N}_{A|C}$ are equal due to the choice of $\Gamma_{AB}=\Gamma_{AC}$. }
\label{WDissPairwise}
\end{figure}

\begin{figure}[h]
\centering
\includegraphics[width=0.8\columnwidth]{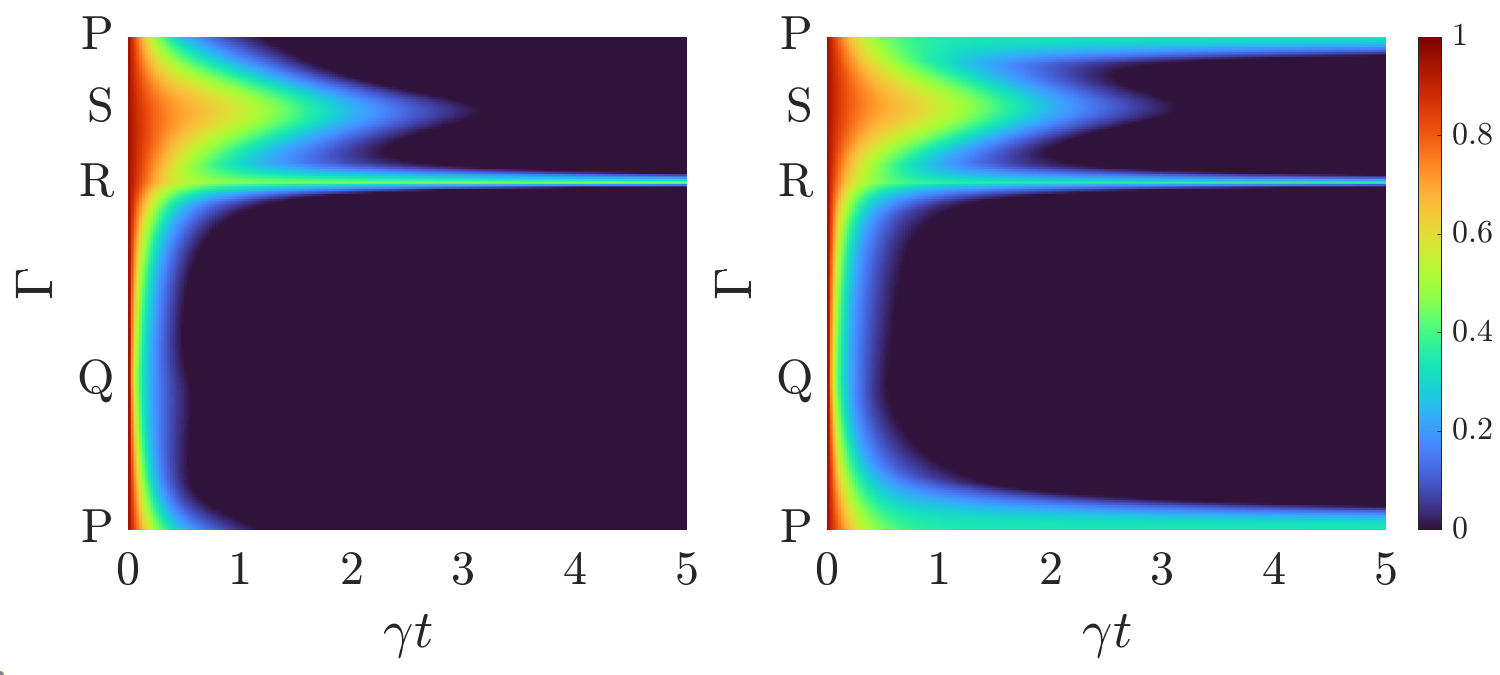}
\caption{Entanglement evolution of bipartions for the W state under amplitude noise along correlation path PQRS for (a) $\mathcal{N}_{A|BC}$ (left) and (b) both $\mathcal{N}_{B|AC}$ and $\mathcal{N}_{C|AB}$ (right). Note that $\mathcal{N}_{B|AC}$ and $\mathcal{N}_{C|AB}$ are equal due to the choice of $\Gamma_{AB}=\Gamma_{AC}$.}
\label{WDissBipartition}
\end{figure}

This mechanism is further illustrated in Figures \ref{WDissPairwise} and \ref{WDissBipartition} which compare the pairwise and bipartition negativities, respectively. At $R$, two pairs of noise environments are maximally anticorrelated, preserving pairwise entanglement between the two related qubit pairs (Fig. \ref{WDissPairwise}a), while the remaining qubit pair under correlated noise disentangles rapidly (Fig. \ref{WDissPairwise}b). Consequently, each bipartition $i|jk$ comprised of at least one preserved entangled pair is then able to evolve toward a steady-state entanglement value. Only at point $R$ do all three bipartitions exhibit comparable steady-state dynamics, yielding a finite long-time tripartite negativity. 

While the W state mainly exhibits this pairwise-to-bipartite entanglement scaffolding, we note a contrasting behaviour at point $S(-\tfrac{1}{2},-\tfrac{1}{2},-\tfrac{1}{2})$, corresponding to equally anticorrelated noise pairs. Here we observe a moderate slowing of tripartite disentanglement at the bipartition level (Fig. \ref{WDissBipartition}) that is not reflected in the pairwise negativities (Fig. \ref{WDissPairwise}). This subtle deviation highlights how multipartite entanglement dynamics under dissipative noise can display behaviors not directly inferable from pairwise dynamics alone, while also confirming the impact of anticorrelated noise on the W state in this case.

\subsection{GHZ-W Superposition States}

We now consider GHZ–W superposition states of the form
\beq
\sqrt{1-p}\,\lvert W \rangle + \sqrt{p}\,\lvert GHZ \rangle, \qquad p \in [0,1]
\eeq
representing a continuous class of states interpolating between the W state ($p=0$) and the GHZ state ($p=1$) \cite{TripartiteNegativity}. Examining the tripartite negativity at a long-time evolution point $t=10\gamma^{-1}$, allows us to identify which mixtures retain non-vanishing entanglement and which undergo full disentanglement over extended timescales.

\begin{figure}[h]
\includegraphics[width=0.49\columnwidth]{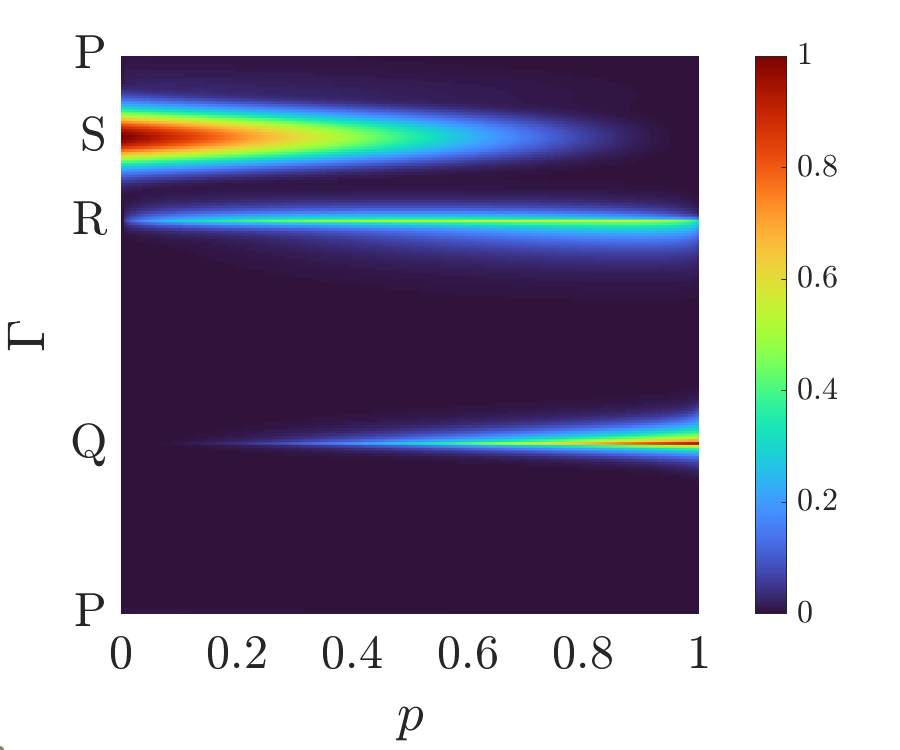}
\hfill    
\includegraphics[width=0.49\columnwidth]{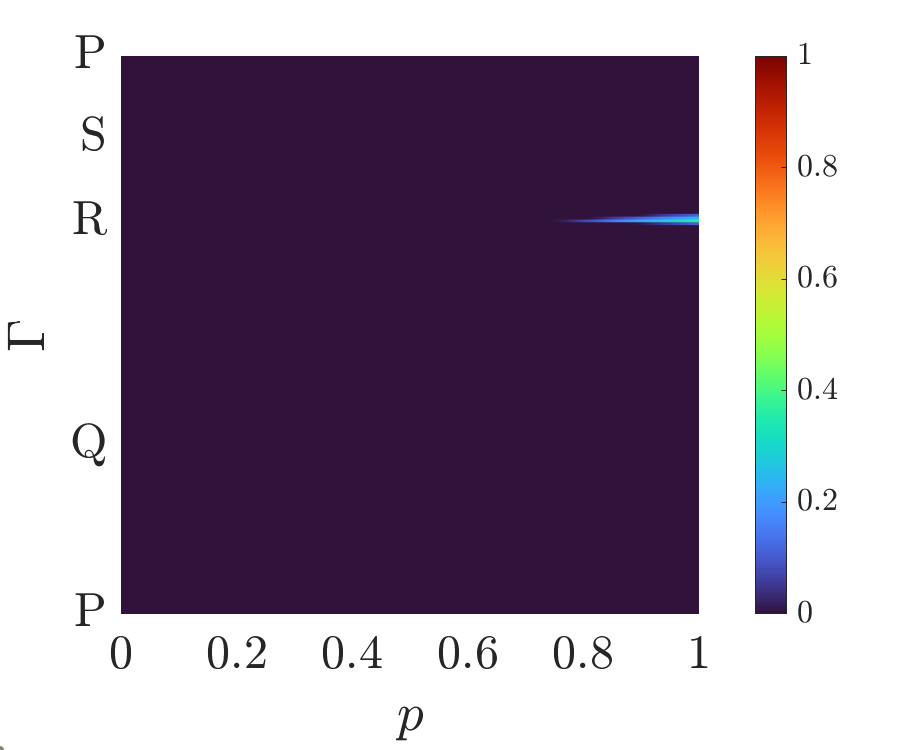}
\caption{$\mathcal{N}_{ABC}$ plotted against $p$ along the path 'PQRS' for mixtures of the W and GHZ states at long timescale $t=10\gamma^{-1}$ under (a) dephasing noise (left) and (b) amplitude noise (right).}
\label{MixedState}
\end{figure}

Figure \ref{MixedState}a illustrates the dependence of the residual tripartite negativity on the weighting parameter $p$ and the noise-correlation configuration. Under dephasing noise, anticorrelated environments at $S(-\tfrac{1}{2},-\tfrac{1}{2},-\tfrac{1}{2})$ most effectively sustain entanglement in GHZ-dominated states ($p\!\to\!0$), whereas fully correlated environments at $Q(1, 1,1)$ favor W-dominated states ($p\!\to\!1$). Notably, for intermediate superpositions ($0<p<1$), partial entanglement persists when two noise environments are anticorrelated and one is correlated, corresponding to $R(-1, -1, 1)$. This persistence is a distinct feature of GHZ-W superpositions and does not appear in either pure-state limit, highlighting the hybrid state's enhanced resilience to certain structured correlations.

Under amplitude noise (Fig. \ref{MixedState}b), the tripartite negativity does not remain fully preserved at long times. However, for sufficiently W-dominated superpositions ($p \gtrapprox 0.75$), a portion of the entanglement survives, approaching a finite asymptotic value that depends on the value of $p$ when two environmental pairs are maximally anticorrelated the the remaining pair is correlated. This residual entanglement originates from the same steady-state behavior identified for the W state at $R(-1, -1, 1)$, and it persists continuously for superpositions that are sufficiently W-dominated. The result highlights that the robustness of GHZ-W superpositions under amplitude noise is directly inherited from the W-state contribution. 

\section{Conclusion}

In summary, we have investigated the dynamics of three-qubit entanglement under correlated noise, analyzing both dephasing and amplitude noise using a Markovian master equation approach. By introducing statistical correlations between pairs of local noise environments, we identified distinct mechanisms through which such correlations influence multipartite entanglement for different initial states.
\par
Under dephasing noise, GHZ and W-type entanglement respond oppositely to the direction of noise correlations. The GHZ state, whose entanglement is intrinsically global, benefits from anticorrelated environments that can, in the ideal limit, completely suppress decoherence at the point $S(-\tfrac{1}{2},-\tfrac{1}{2},-\tfrac{1}{2})$. In contrast, the W state, which retains bipartite correlations in its reduced subsystems, is stabilized by correlated environments, achieving full preservation at $Q(1, 1, 1)$. These contrasting dependencies arise from the additive versus pairwise (product-like) roles of the correlation parameters in the respective tripartite negativities, giving us a direct view into their fundamentally different underlying entanglement structures.
\par
Under amplitude noise, both GHZ and W states experience entanglement sudden death; however, correlations can slow this process. For the GHZ state, specific correlation configurations delay disentanglement but cannot prevent it. The W state displays more nuanced behavior. At $R(-1, -1, 1)$, the tripartite negativity asymptotically approaches a steady-state value due to the persistence of pairwise entanglement within two anticorrelated subsystems. This residual entanglement reveals that appropriately structured noise correlations can stabilize parts of multipartite entanglement, even under dissipative dynamics.
\par
Extending to GHZ-W superposition states, we observed a continuous transition between these regimes. Under dephasing noise, entanglement protection shifts smoothly from the GHZ-dominated regime favoring correlated noise to the W-dominated regime favoring anticorrelated noise. Additionally, we observed entanglement partially preserved for intermediate states at point $R(-1, -1, 1)$ that did not appear in the W or GHZ state limits. Under amplitude noise, the steady-state behavior characteristic of the W state persists for sufficiently W-dominated superpositions, maintaining a finite asymptotic entanglement that inherits directly from the W component.
\par
Together, these results elucidate how noise correlations can influence and, in certain regimes, stabilize multipartite entanglement. Beyond their theoretical relevance, this framework provides intuition for how different entanglement structures respond to correlated environments and may guide future considerations for noise-prone multi-qubit quantum architectures.

\bibliography{QIP-3QMarkov-Refs}

\end{document}